# Interactions between a propagating detonation wave and water spray cloud in hydrogen/air mixture

Yong Xu, Huangwei Zhang*

*Department of Mechanical Engineering, National University of Singapore, 9 Engineering Drive 1, Singapore, 117576, Republic of Singapore*

**Abstract**

Inhibition of hydrogen explosion is crucial to realize its wide applications and fine water spray is an ideal mitigant due to numerous advantages. In this work, interactions between a propagating hydrogen/air detonation wave and circular water cloud are numerically studied. Eulerian−Lagrangian method involving two-way gas-droplets coupling is applied, with a two-dimensional configuration. Different droplet (diameter, concentration) and cloud (diameter) properties are considered. Our results show that droplet size, concentration and cloud radius have significant effects on peak pressure trajectory of the detonation wave. After interacting with cloud, the detonation wave exhibits three propagation modes, including perturbed propagation, leeward re-detonation, and detonation extinction. Leeward re-detonation is analyzed from unsteady evolutions of gas and liquid droplet quantities. The refracted detonation wave inside the cloud is decoupled and propagates more slowly than the one outside the cloud. The detonation re-initiation is from a local hot spot, caused by shock focusing from upper and lower diffracted detonations. Disintegration of water droplets proceeds when the detonation wave crosses the cloud and multiphase interfacial instability is observed due to the difference in effective density of the two fluids. Furthermore, detonation extinction is observed when we consider various water cloud size. It is featured by quickly fading peak pressure trajectories when the detonation passes the cloud, and no local autoignition occurs in the shock focusing area. Evolutions of thermochemical structures from the shocked area in an extinction process are also studied. Moreover, parametric studies considering various droplet concentrations and cloud radii are performed. It is shown that the critical cloud size to quench a detonation decreases when the droplet concentration is increased. However, when the droplet concentration is beyond 0.84 kg/m$^3$, the critical cloud size is negligibly influenced.

*Corresponding author. Email:
huangwei.zhang@nus.edu.sg.

## 1. Introduction

Hydrogen ($H_2$) as a clean fuel has been deemed a promising solution to reducing carbon emissions. Nonetheless, $H_2$ leakage, e.g., from pressurized tanks, may cause ignition and even detonation, due to its low ignition energy and wide flammability limit. Therefore, inhibition of $H_2$ explosion hazards is crucial to its practical applications in the foreseeable future. Water spray with fine droplets is an ideal mitigant for gas explosions [1], because it can absorb heat from gas phase due to large heat capacity, latent heat of evaporation and specific surface area.

Numerous studies about shock or blast attenuation by water sprays have been reported, as summarized in Ref. [2]. For instance, Jourdan et al. [3] use water aerosol shock tube experiments to study shock attenuation in a cloud of water droplets. They correlate the shock attenuation degree with shock tube and droplet properties. Using the similar facilities, Chauvin et al. [4] find the peculiar pressure evolution after the transmitted shock in two-phase mixture and they also measure the overpressures under different water spray conditions. Adiga et al. [5] unveil the physical picture of fine water droplet breakup upon shock passage and quantify the droplet fragmentation with breakup energy. Eulerian − Lagrangian simulations are performed by Ananth et al. [6] to examine the effects of mono-sized fine water mists on a confined blast. It is concluded that the latent heat absorption is dominant for blast mitigation, followed by convective heat transfer and momentum exchange. Schwer and Kailasanath [7] simulate unconfined explosions in water sprays and find that momentum extraction by water droplets plays a more important role in weakening the blast than droplet evaporation.

There are also studies about how water sprays affect detonation propagation. Thomas et al. [8], for example, attribute detonation failure to high heat loss by water droplets. The water droplet diameter and loading are identified as key factors for detonation inhibition. Niedzielska et al. [9] also observe that small droplets have strong influence on quenching a detonation due to their fast evaporation rate. Jarsalé et al. [10] find that water droplets do not alter the ratio of hydrodynamic thickness to detonation cell size, but can influence the detonation stability. Besides the foregoing experimental work, Watanabe et al. [11] observe from their simulations that the cellular structures of hydrogen detonations with water sprays become more regular, compared to the droplet-free detonations. Their results also show that droplet breakup mainly occurs near the shock front, and the average diameter of the disintegrated water droplets is independent on the initial propagation velocity of the shock front [12].

Furthermore, propagation regimes of hydrogen/air detonations in water mists are predicted in the parameter space of droplet loading and diameter by Xu et al. [2]. They also reveal the evolutions of the characteristic chemical structure and timescale in the induction zone for both propagating and failed detonations. In addition, it is also seen that the interphase energy and momentum transfer have more direct influence on induction zone than water droplet evaporation. More recently, Xu and Zhang [13] study the unsteady phenomena in hydrogen/oxygen/argon detonation with water mists. They find that existence of water mists results in detonation galloping motion and even failure, but these effects depend on droplet properties. In most preceding studies (e.g., [8]-[13]), sprayed water is considered as a flooding species in the entire studied domain. However, such idealized distributions would be practically rare, considering relatively long droplet dispersion timescale compared to that of detonation wave (DW). Instead, water droplets for explosion suppression are largely generated from properly arranged sprinklers, leading to localized water spray cloud. How a DW interacts with a water cloud and the performance of water cloud for detonation inhibition have not been explored yet.

In this work, hydrogen/air detonations interacting with a circular water spray cloud are simulated. The Eulerian-Lagrangian method considering two-way gas−liquid coupling will be used. The effects of water droplet properties and cloud size on incident detonations is studied. The objectives of this work are to clarify: (1) the effects of water cloud on a propagating detonation wave; (2) evolution and mechanism of hydrogen re-detonation at the leeward side of a cloud; and (3) detonation extinction by a water cloud. The results from this work will be useful to provide scientific evidence for mechanism and performance of hydrogen detonation inhibition with water sprays in practical applications.

## 2. Mathematical model

Interactions between an incident detonation wave and circular water spray cloud will be studied. The Eulerian−Lagrangian method is used to model the two-phase, multi-species, compressible, reacting flows. For the gas phase, the Navier-Stokes equations of mass, momentum, energy, and species mass fraction are solved. For the liquid spray phase, the Lagrangian method is employed to track the individual droplets. Droplet collisions are neglected because dilute sprays are considered (initial droplet volume fraction < 0.1%). Droplet breakup by aerodynamic force is modelled following Pilch and Erdman [15]. It is also assumed that the temperature inside the droplets is uniform due to their small Biot numbers (< 0.0013). Gravitational force is not considered due to smallness of the droplets. Therefore, the equations of mass, velocity and temperature of a single droplet read

$$\frac{dm_d}{dt} = -\dot{m}_d, \quad (1)$$

$$m_d \frac{d\mathbf{u}_d}{dt} = \mathbf{F}_p + \mathbf{F}_d, \quad (2)$$

$$c_{p,d} m_d \frac{dT_d}{dt} = \dot{Q}_c + \dot{Q}_{lat}, \quad (3)$$

where $t$ is time, $m_d$ is the mass of a single droplet, $\dot{m}_d$ is the droplet evaporation rate, modelled as $\dot{m}_d = k_c A_d W_d (c_s - c_g)$. $A_d$ is the surface area of a single droplet, $k_c$ the mass transfer coefficient, and $W_d$ the molecular weight of the vapor. $c_S$ and $c_g$ are the vapor mass concentrations at the droplet surface and in the gas phase, respectively. In Eq. (2), $\mathbf{u}_d$ is the droplet velocity vector. The pressure gradient force $\mathbf{F}_p$ is calculated as $\mathbf{F}_p = -V_d \nabla p$, in which $V_d$ is the volume of a droplet. The drag force $\mathbf{F}_\mathrm{d}$ is $\mathbf{F}_d = (18\mu/\rho_d d_d^2) \cdot (C_d Re_d/24) \cdot m_d (\mathbf{u} - \mathbf{u}_d)$. $Re_d$ is the droplet Reynolds number, $C_d$ the drag coefficient (modelled following Schiller and Naumann [16]), $\mu$ the gas dynamic viscosity, $\rho_d$ the water material density, $d_d$ the droplet diameter, and $\mathbf{u}$ the gas velocity vector at the droplet location. In Eq. (3), $c_{p,d}$ is the droplet heat capacity, and $T_d$ is the droplet temperature. The convective heat transfer rate $\dot{Q}_c = h_c A_d (T - T_d)$, where $T$ is the gas temperature and $h_c$ is the convective heat transfer coefficient, estimated with Ranz and Marshall correlations [17]. $\dot{Q}_{lat}$ is the latent heat transfer rate by droplet evaporation. Two-way coupling between gas and droplets are implemented for mass, momentum, and energy exchanges.

The gas and liquid phase equations are solved using an OpenFOAM solver *RYrhoCentralFoam* [18-20]. Detailed validations and verifications have been done previously against experimental and/or theoretical data [18], including shock capturing, shock-eddy interaction, molecular diffusion, flame-chemistry interactions, single droplet evaporation, and two-phase coupling. For the gas phase equations, second-order backward scheme is employed for temporal discretization and the time step is about $7\times10^{-10}$ s. A MUSCL-type scheme with van Leer limiter is used for convective flux calculations in momentum equations. Total variation diminishing scheme is applied for the convection terms in energy and species equations. Also, second-order central differencing is applied for the diffusion terms. A mechanism with 13 species and 27 reactions [21] is used for hydrogen combustion. For the liquid phase, Eqs. (1)-(3) are integrated with a Euler implicit method and the right terms are treated with a semi-implicit approach. Details about the numerical method can be found in Refs. [2,18].

## 3. Physical model and numerical setup

The schematic of the physical problem is shown in Fig. 1. The length ($L$, $x$-direction) and width ($W$) of the computational domain are 0.3 m and 0.025 m, respectively. It includes a driver section ($x$ = 0 – 0.19 m, not shown in Fig. 1) and detonation – cloud interaction section ($x$ = 0.19 – 0.3 m). One circular cloud is beforehand placed in the second section, to mimic the water sprays generated from a sprinkler to inhibit explosion accident. The cloud contains many water droplets, and the actual droplet number is determined from the initial droplet diameter $d_d^0$ and droplet concentration $c$. The cloud diameter is aligned with $y$ = 0 (termed as "centerline" hereafter). The cloud leftmost point (see Fig. 1) always lies at (0.19375 m, 0 m); instead, the cloud center, ($x_c$, 0), is varied when the cloud radius $R$ is changed. Therefore, a water cloud is parameterized by its geometry ($R$ or $x_c$) and droplet properties ($d_d^0$ and $c$).

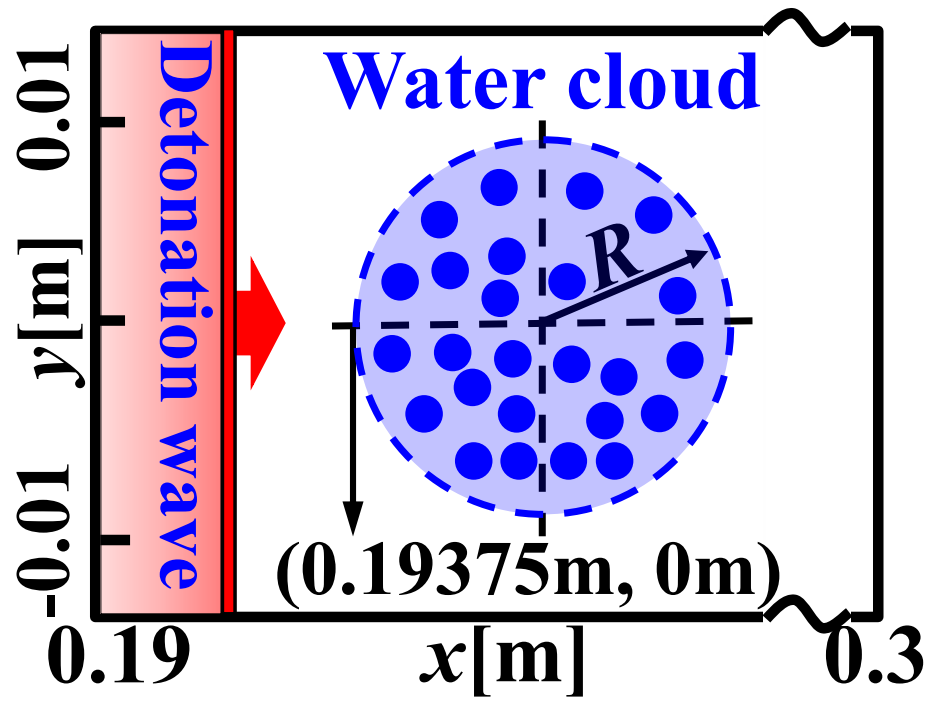

Fig. 1. Detonation-cloud interactions. $R$: cloud radius.

Initially ($t$ = 0), the entire domain is filled with stoichiometric $H_2$/air mixture, with temperature and pressure being $T_0$ = 300 K and $p_0$ = 50 kPa, respectively. The detonation is initiated by three vertically placed hot spots (2,000 K and 5 MPa) near the left end ($x$ = 0 m). Inside the cloud, mono-sized water droplets (diameter $d_d^0$ = 1 – 15 μm, and concentration $c$ = 0.105 – 1.68 kg/m$^3$) are uniformly distributed. They are static ($\mathbf{u}_d^0$ = 0) before the detonation arrives. Their initial temperature, material density and heat capacity are 300 K, 997 kg/m$^3$ and 4,187 J/kg·K, respectively. The radius of the water cloud, $R$, varies from 0.00625 to 0.0125 m, i.e., 25%-50% of the domain width $W$.

The upper and lower boundaries of the domain in Fig. 1 are periodic. For the left boundary ($x$ = 0), the non-reflective condition is enforced for the pressure, while the zero-gradient condition applied for other quantities. For the boundary at $x$ = 0.3 m, zero gradient is assumed for all variables.

The domain in Fig. 1 is discretized with uniform Cartesian cells of 20×20 μm$^2$. The half-reaction length in the ZND structure of stoichiometric $H_2$/air detonation under the current investigated conditions is approximately 382 μm. Considering the lengthened induction zone of spray detonations due to the evaporating droplets [2,11], more than 19 cells are in the induction zone in the simulations. Mesh sensitivity is also studied, showing that the detonation cellular structure is almost not affected when the mesh size is halved (see supplementary document). Moreover, with current Eulerian mesh resolution and diameter of Lagrangian droplets, an error analysis for droplet evaporation timescale considering the effects of Pélect number [22] is presented in supplementary document. It is shown that the relative error of droplet evaporation time is below 10% in this work.

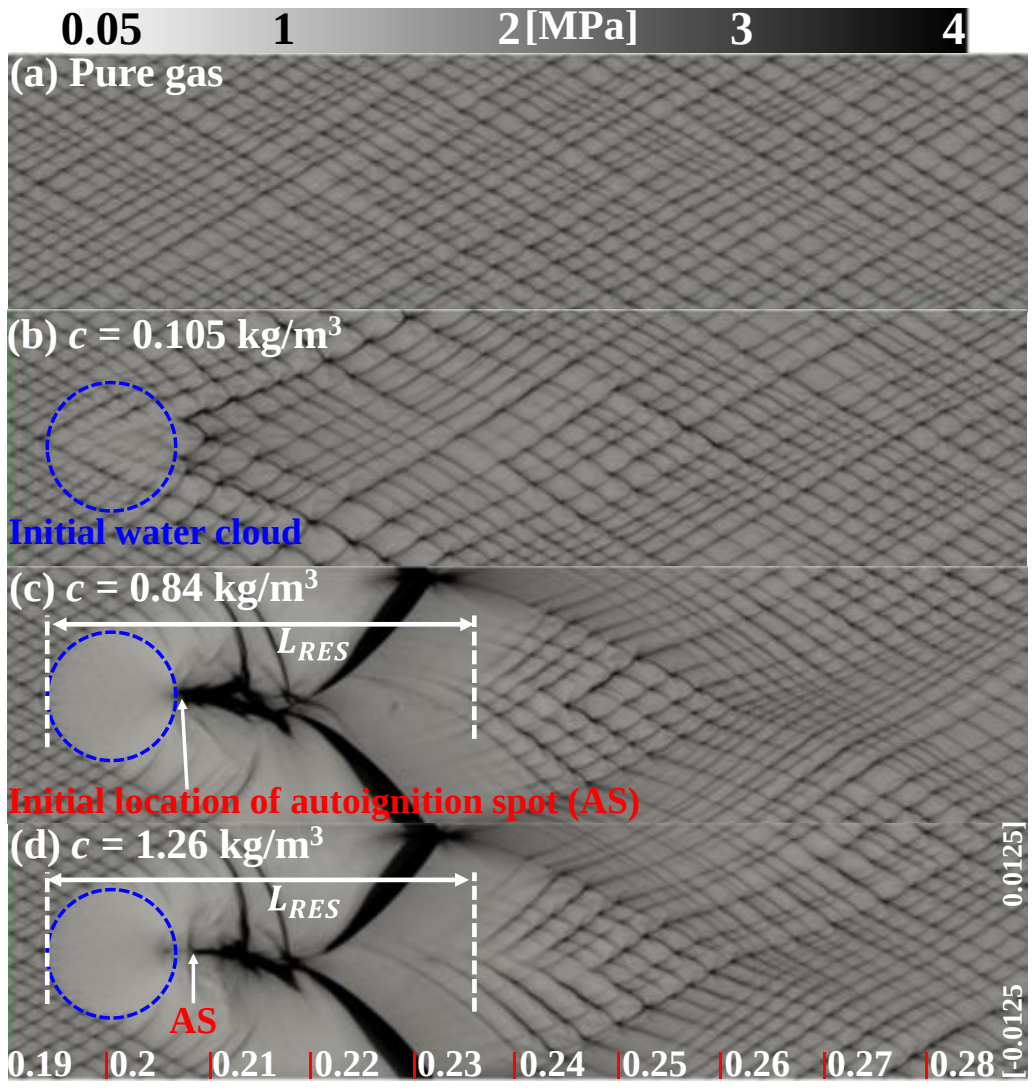

Fig. 2. Trajectories of peak pressure with various droplet concentrations. $d_d^0$= 2.5 μm, $R$ = 0.25$W$. Axis label unit: m.

## 4. Results and discussion

### *4.1 Water droplet size and concentration effects*

Figure 2 shows the trajectories of peak pressure in detonation-cloud interaction section when different droplet concentrations $c$ are considered. The initial droplet diameter is 2.5 μm and initial cloud radius is $R$ = 0.25 $W$. The cloud-free ($c$ = 0) results are shown in Fig. 2(a). Evident from Fig. 2 is that the cellular structures change, with different degrees, after the DWs cross the water cloud (their initial positions are marked as blue circles, and their movement will be discussed in Section 4.2). For low droplet concentration, e.g., $c$ = 0.105 kg/m$^3$ in Fig. 2(b), the averaged cell width behind the water cloud is slightly increased, indicating the enhanced frontal instability in the post-cloud area. Nonetheless, the cell size and regularity are quickly restored when $x$ > 0.23 m. We term this mode as *perturbed propagation*. Furthermore, when $c$ ≥ 0.84 kg/m$^3$, maximum pressures are immediately reduced to below 1 MPa in the cloud, as found from Figs. 2(c) and 2(d). This means that the gas reactivity at the triple points is significantly reduced because of the high-concentration water mists. Actually, the DWs experience localized extinction with decoupled shock front (SF) and reaction front (RF). Behind the cloud, the peak pressure trajectories generally become blurred. However, a horizontal black strip with high overpressure approximately extends from the rightmost point of the cloud, which corresponds to the evolution of an igniting hot spot (labelled with AS) due to shock compression. In Figs. 2(c) and 2(d), at about $x$ = 0.22 m, the strip bifurcates into two thick trajectories, which are the upper and lower transverse detonation connected with a Mach stem generated from a re-initiation process. We term this process as *leeward re-detonation*, and the underlying mechanism will be discussed in Section 4.2. At $x$ = 0.24 m, cellular structures appear again. The mean cell widths are close (about 1.9 mm) in Figs. 2(c) and 2(d), similar to that (1.7 mm) before the DW interacts with the water cloud. The distance between the cloud leftmost point and minimum $x$ coordinate with clear cellular structure can be defined as DW restoration distance, which is $L_{RES}$ ≈ 41 mm in Figs. 2(c) and 2(d).

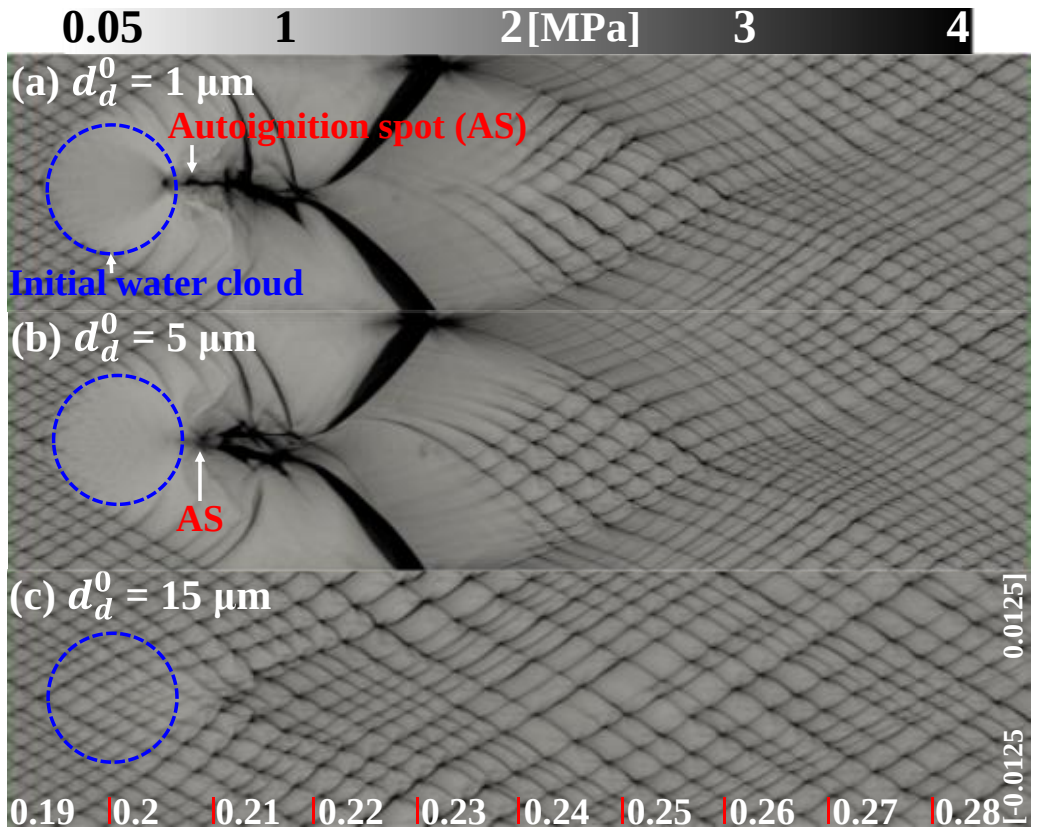

Fig. 3. Trajectories of peak pressure with various droplet diameters. $c$ = 0.84 kg/m$^3$, $R$ = 0.25$W$. Axis label unit: m.

Figure 3 presents the peak pressure trajectories with varying droplet diameters, $d_d^0$ = 1, 5, and 15 μm. Here $c$ = 0.84 kg/m$^3$ and $R$ = 0.25 $W$. Note that the results of 2.5 μm are already shown in Fig. 2(c). For small droplets (i.e., 1 and 5 μm in Figs. 3a and 3b), the DWs experience leeward re-detonation process and the evolutions of cellular structures are like what have been discussed in Fig. 2(c). However, for large droplets in Fig. 3(c), DW extinction does not occur because of slower exchange rate for mass and energy; instead, perturbed DW propagation occurs. Specifically, the detonation cells become more irregular, and the averaged cell width is increased, thereby more unstable front after it crosses the cloud.

In both perturbed propagation and leeward detonation modes, detonation-cloud interactions can be further quantified by lead shock propagation speed in the cloud, $D$, which is calculated from the cloud diameter ($2R$) divided by the time duration with which the shock respectively crosses the leading and trailing points along the centerline ($y$ = 0). Plotted in Fig. 4 are the effects of droplet concentration and diameter on $D$ (scaled by the C-J speed of droplet-free $H_2$/air mixture, $D_{CJ}$ = 1,961 m/s) when $R$ = 0.25 $W$. Note that the speed in the cloud-free case is calculated based on the time for the DW to pass the leftmost and rightmost edges of the cloud, although the cloud is not placed. One can see that perturbed propagation (solid symbols) only occurs when droplet size is large and $c$ is small. Generally, $D$ is below that of the cloud-free case. For a constant droplet concertation (e.g., 0.84 kg/m$^3$), $D$ decreases when the droplet size becomes smaller. This is because the smaller diameter corresponds to a larger specific surface area, and therefore faster interphase exchange of mass, momentum, and energy. Moreover, when the droplet diameter is fixed (e.g., 10 μm), $D$ decreases with $c$. This is caused by the enhanced influences of fine droplets on the DW due to increased droplet concentration.

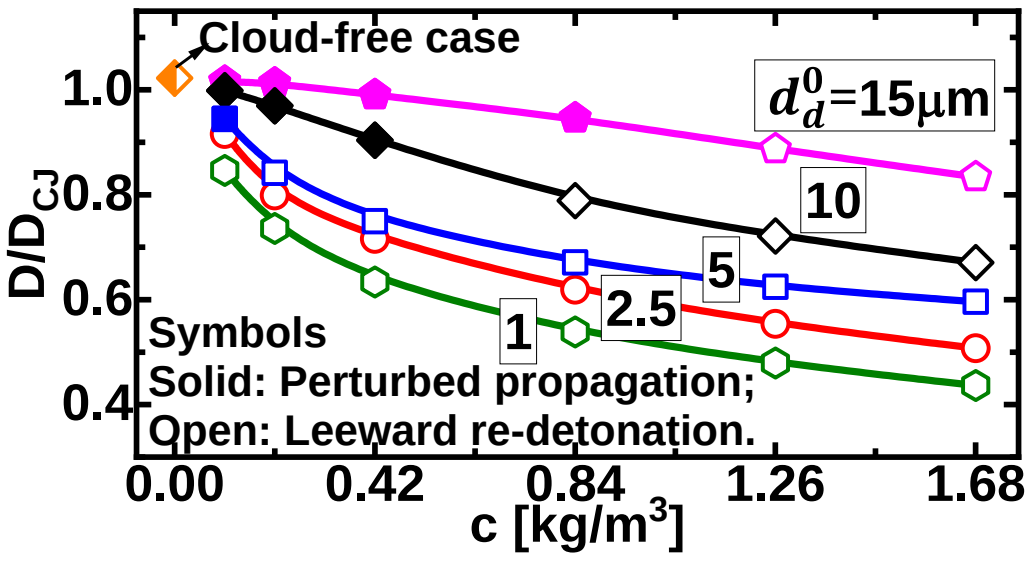

Fig. 4. Average shock speed for crossing water cloud as a function of droplet concentration and diameter. $R$ = 0.25$W$.

### *4.2 Leeward re-detonation*

Figure 5 shows a leeward re-detonation process when $c$ = 0.84 kg/m$^3$, $d_d^0$= 2.5 μm, and $R$ = 0.25 $W$. This corresponds to Fig. 2(c). At 98 μs in Fig. 5(a), the temperature of the refracted DW inside the cloud (termed as inner DW) is reduced to around 2,150 K. At 102 μs, the RF and SF inside the cloud decouple. The RF is deemed quenched because of lower temperature (~1,600 K) and HRR (~1×10$^{12}$ J/m$^3$/s) around it. However, beyond the cloud, the detonation wave (termed as outer DW) remains. It starts to diffract after passing the southern/northern poles of the cloud and propagates relatively fast compared to the quenched one in the cloud. The outer (diffracted, burning) and inner (refracted, quenched) sections are connected at the cloud boundary (dashed line). From 104 to 107 μs, the outer DW becomes gradually convex with respect to the fresh gas, and the upper / lower RFs move gradually closer. Meanwhile, decoupling of SF and RF also occur for the outer DW near the cloud. This is associated with the reduced intensity of lead shock in the outer DW, due to the change of the diffracted shock curvature.

In Fig. 5(e), shock focusing occurs: the SFs degraded from the upper and lower parts of the outer DW superimpose near the centerline. This leads to high overpressure (> 3 MPa) and temperature (> 1,600 K) at the point S1. Meanwhile, two peninsula-shaped RFs from the upper and lower DWs are approaching each other near that point. At 108.5 μs, an isolated autoigniting spot (AS) arises, which generates higher pressure and temperature there, indicating the onset of thermal runaway in an isochoric combustion. Note that now the refracted shock has left the cloud (i.e., AS is at the cloud leeward). We compile the AS locations from all re-detonation cases in Fig. 4 (see supplementary document) and it is found that AS occurs almost along the centerline and their locations ($x$ coordinate) range from 2.1$R$ to 3.3$R$, relative to the leading point of the cloud. This corresponds to external shock focusing [23], which is essentially determined by the differentiated timescales for diffracted and refracted shocks [24]. Subsequently, AS quickly develops into detonation waves at 111 μs in Fig. 5(h). Their transverse propagation consumes the gas between the lead SF and RF. Meanwhile, it also overtakes the lead SF and an overdriven Mach stem is formed in Fig. 6(i). The upward- and downward-running transverse DWs generate the thick bifurcated trajectories in Fig. 2(c).

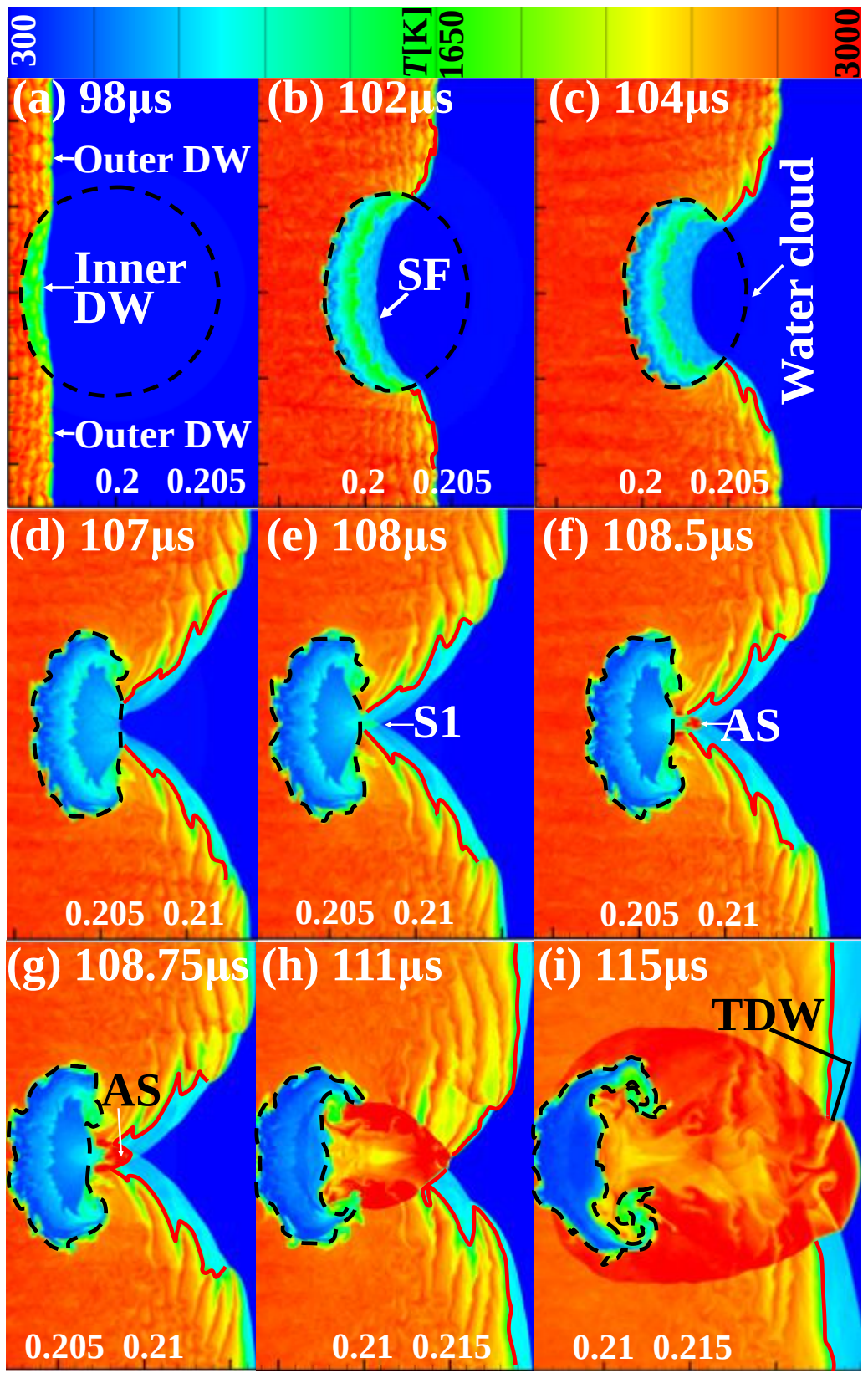

Fig. 5. Evolutions of gas temperature in re-detonation. $c$ = 0.84 kg/m$^3$, $d_d^0$= 2.5 μm, $R$ = 0.25$W$. Axis label unit: m. Dashed lines: cloud boundary; RF/SF: reaction/shock fronts; AS: auto-ignition spot; TDW: transverse detonation wave.

To further describe the mechanism of leeward re-detonation, chemical explosive mode analysis [25] is performed to extract the characteristic chemical information. Figure 6 shows the spatial profiles of pressure, HRR, and eigen values $\lambda_e$ of the chemical Jacobian along the centerline at four instants. For clear illustration, the signed exponent of the eigenvalue $\lambda_e$ of chemical explosive mode is plotted, i.e., $\lambda_{CEM} = sign[Re(\lambda_e)] \cdot log_{10}[1 + |Re(\lambda_e)|]$ . Positive (negative) $\lambda_e$ indicates the propensity of chemical explosion (non-reactive or burned mixtures). In Fig. 6(a), the RF and SF are immersed in the water cloud (shaded area). Despite detonation extinction, the mixture in the induction zone between RF and SF still has explosion propensity with high $\lambda_{CEM}$. Meanwhile, the pressure behind the SF is slightly increased because of shock compression. The water vapor mass fraction gradually increases towards the RF. At 108 μs, $\lambda_{CEM}$ peaks behind the SF due to shock focusing, evidenced by the elevated pressure, HRR peaks and high $\lambda_{CEM}$ at the same location. At $x$ = 0.204-0.205 m, the mixture reactivity is low (smaller $\lambda_{CEM}$) because of droplet evaporation and/or interphase heat transfer inside the cloud. This can also be confirmed by finite water vapor mass fraction $Y_{H2O}$ (Fig. S4 of the supplementary document) ahead of the RF. In Fig. 6(c), the autoigniting spot (AS in Fig. 5f) is generated, corresponding to locally negative $\lambda_{CEM}$ and two RFs with high HRR (> $4\times10^{13}$ J/m$^3$/s) at $x$ = 0.2076 and 0.2086 m. For the same reason, pressure is also increased near the AS location in Figs. 6(c) and 6(d). Moreover, at 108.75 μs, the right RF is coupled with the SF, generating a new detonation wave (i.e., evolving into the Mach stem in Fig. 5i).

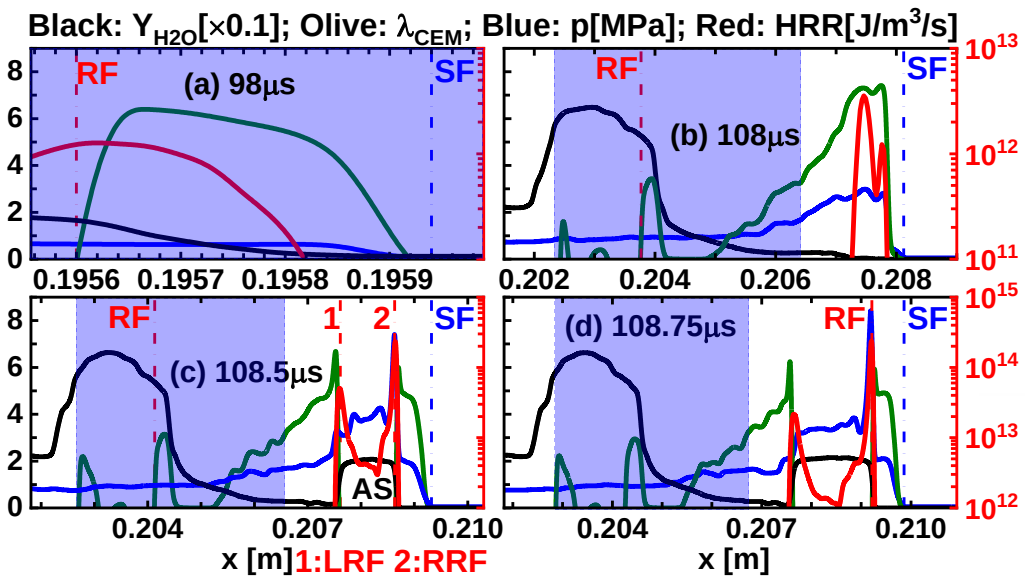

Fig. 6. Evolutions of $\lambda_{CEM}$, pressure, HRR, and water vapor mass fraction around the autoigniting spot. $c$ = 0.84 kg/m$^3$, $d_d^0$= 2.5 μm, and $R$ = 0.25$W$. Shaded area: water cloud.

Time sequence of the Lagrangian water droplets (colored with the diameter) in Fig. 5 are presented in Fig. 7. At 102 μs, the water droplets start to break up when the DW encroaches. A delay (about 3 mm) can be found between the refracted DW and disintegrated droplets near the left cloud boundary. The droplet size is reduced to below 1.5 μm from its initial value (2.5 μm). At 108.5 μs, the shock just leaves the cloud and most of droplets are highly disintegrated, and the number of child droplets increases quickly. Some jets of fine droplets (below 1 μm) appear near the southern/northern poles of the cloud, caused by the vortex shedding due to the impulsive shock acceleration of a perturbed interface between multiphase fluids [26,27]. In Fig. 7(c), a jellyfish-shaped cloud appears, with two tails of fine droplets extending from the shrinking cloud. The instability of the cloud boundary is evaluated by an effective Atwood number [27], i.e., $A_e \equiv (\rho_{e2} - \rho_{e1})/(\rho_{e2} + \rho_{e1})$, where $\rho_{e1}$ and $\rho_{e2}$ are the effective densities of the background gas and the two-phase fluid, respectively. When $A_e$ is close to 0, the interfacial fluids develop into fingers-like structures; For $A_e$ close to 1, the light fluid become "bubbles" below the heavy fluid. In this case, $A_e$ is approximately 0.5, and the morphology of interfacial instability is complicated. Expanded discussion is beyond the scope of this work, and it merits a further study.

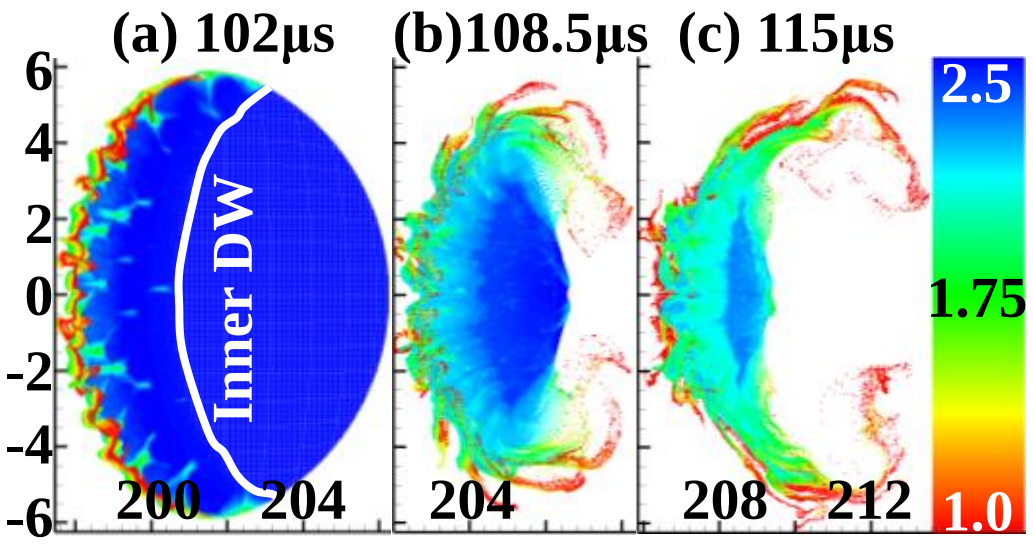

Fig. 7. Evolutions of Lagrangian droplets colored by diameter (in μm). $c$ = 0.84 kg/m$^3$, $d_d^0$= 2.5 μm, and $R$ = 0.25$W$. Axis label unit: mm.

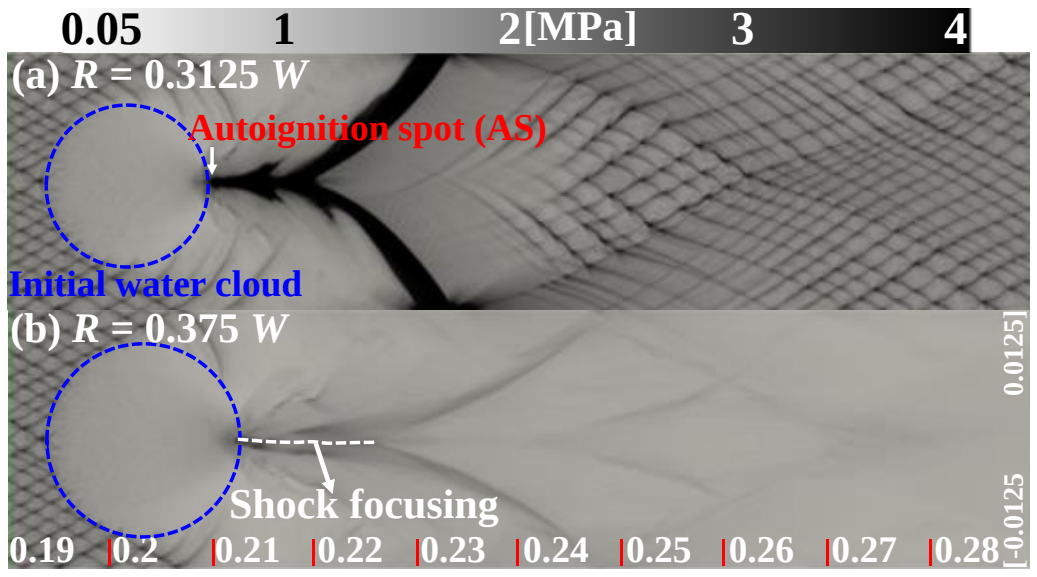

Fig. 8. Numerical soot foils change with various cloud sizes. $c$ = 0.84 kg/m$^3$ and $d_d^0$= 2.5 μm. Axis label unit: m.

### 4.3 *Detonation extinction*

Up to this point, the circular cloud radius is fixed to $R$ = 0.25$W$. The effects of cloud size on the incident detonation wave will be studied in this section, through the peak pressure trajectories in Fig. 8. Two additional cloud radii, $R$ = 0.3125$W$ and 0.375$W$, are considered with fixed $c$ = 0.84 kg/m$^3$ and $d_d^0$= 2.5 μm. For the cloud size in Fig. 8(a), the above-mentioned extinction / re-detonation process in Fig. 2(c) with $R$ = 0.25 $W$ is observed. However, in the larger cloud in Fig. 8(b), the incident DW is quenched after crossing it, characterized by quickly fading peak pressure trajectories, and no re-initiation is observed. The transient of DW extinction in Fig. 8(b) is visualized in Fig. 9. The detonation frontal structures in Figs. 9(a)-9(d) share the similar evolutions with the Fig. 5. The outer DW propagates faster than the inner one, leading to shock focusing indicated by S1 in Fig. 9(e). However, in this scenario, no autoigniting spot is observed, and the distance between the lead SF and RF is gradually enlarged, as shown in Fig. 9(f). The intrinsic chemistry of the focused area will be discussed in Fig. 10.

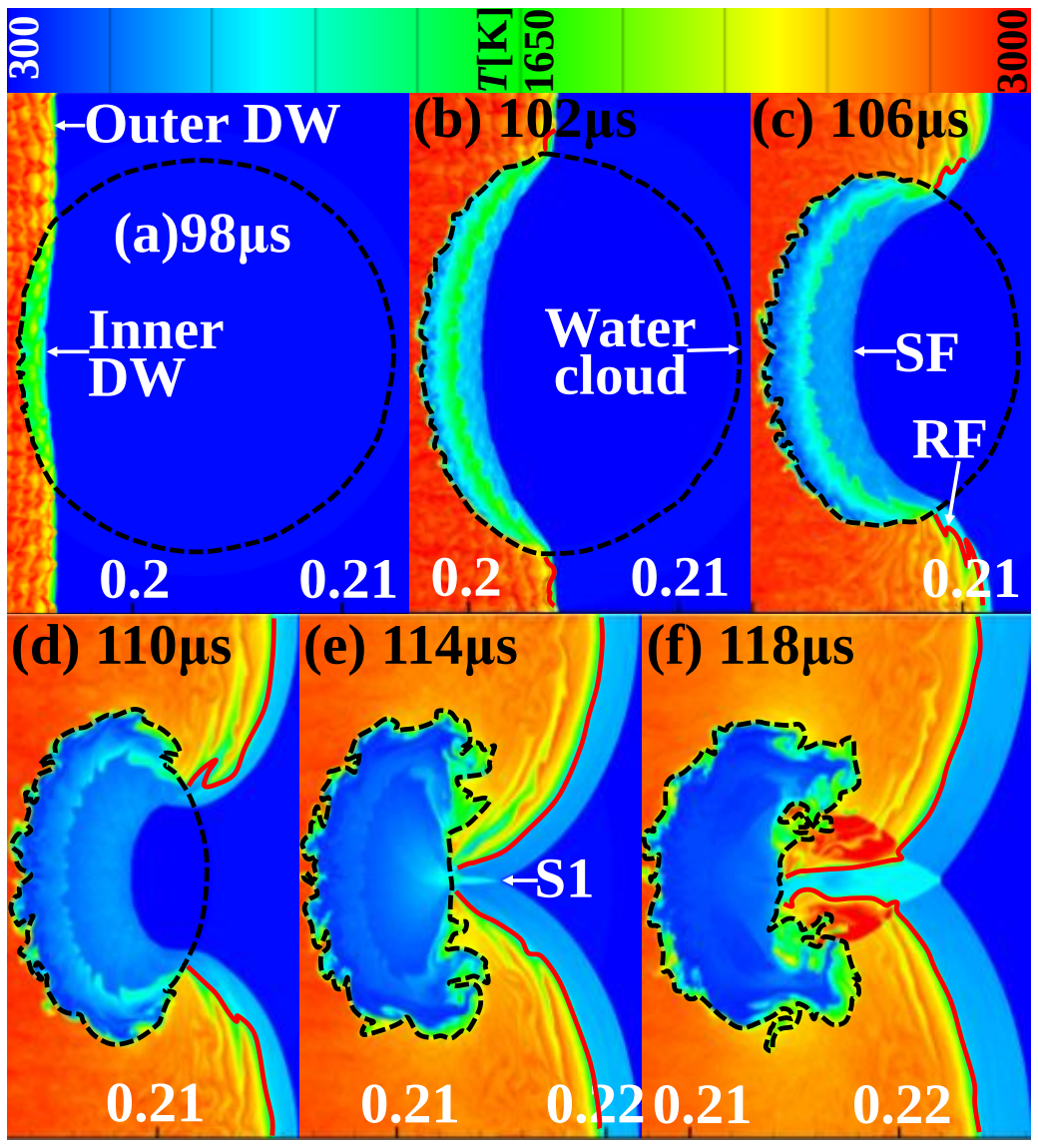

Fig. 9. Time history of gas temperature in a quenched detonation. $c$ = 0.84 kg/m$^3$, $d_d^0$= 2.5 μm, and $R$ = 0.375 $W$. S1: shock focusing location. Axis label unit: m.

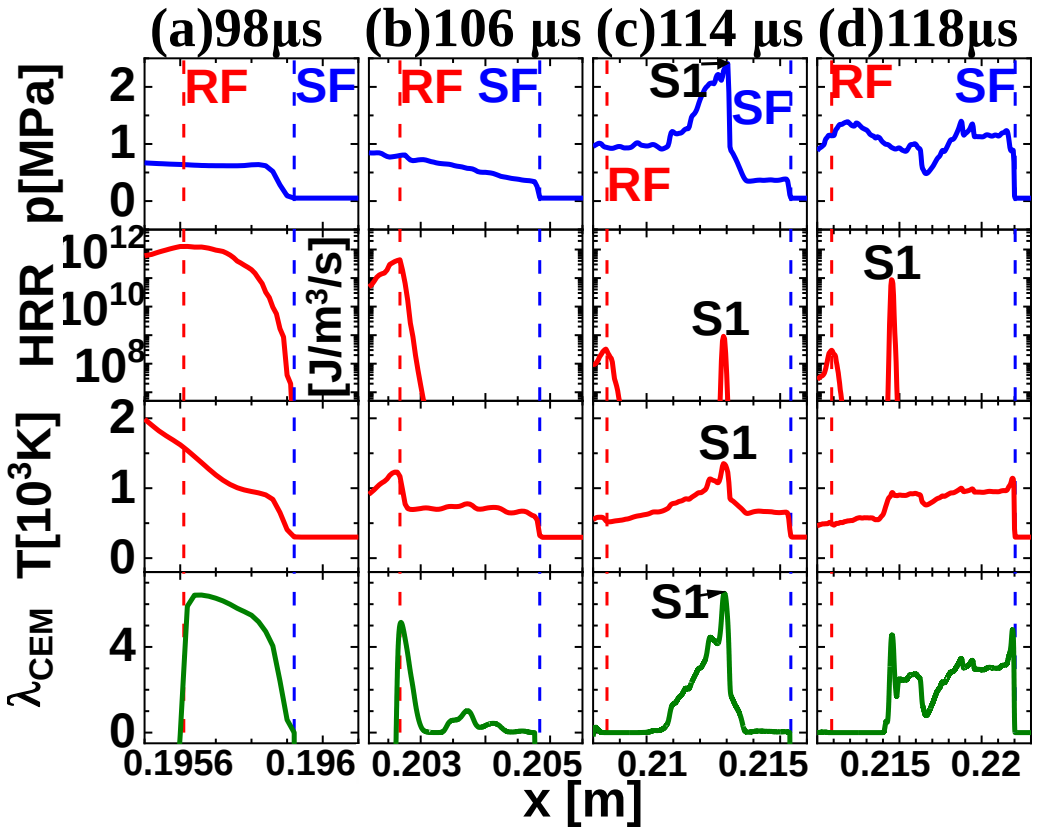

Fig. 10. Evolutions of thermochemical structures along the centerline between the RF and SF. $c$ = 0.84 kg/m$^3$, $d_d^0$= 2.5 μm, and $R$ = 0.375$W$.

Time history of pressure, HRR, gas temperature, and $\lambda_{CEM}$ are plotted in Fig. 10. They are extracted along the domain centerline in Fig. 9. As mentioned above, no autoignition at the shock focusing location (S1) is observed. The peak values of pressure, temperature, HRR and $\lambda_{CEM}$ at S1 in Fig. 10(c) is lower than the counterpart of S1 in Fig. 6(b). The differentiated thermochemical conditions in the two autoignition spots can be found from Table S1 of the supplementary document. Specifically, peak pressure (2.4 MPa) and temperature (1,344 K) of re-initiation failure are much lower than that the successful one (3 MPa and 1,602 K, respectively). This cannot create an AS to re-initiate a detonation in Fig. 10(c), although the shocked gas mixture is highly chemically explosive around S1 ($\lambda_{CEM}$ = 6.5). This can be further confirmed by the reduction of pressure, temperature and $\lambda_{CEM}$ around S1 in Fig. 10(d). It is also demonstrated by almost unconsumed $H_2$/$O_2$ around S1 (Fig. S5 in supplementary document).

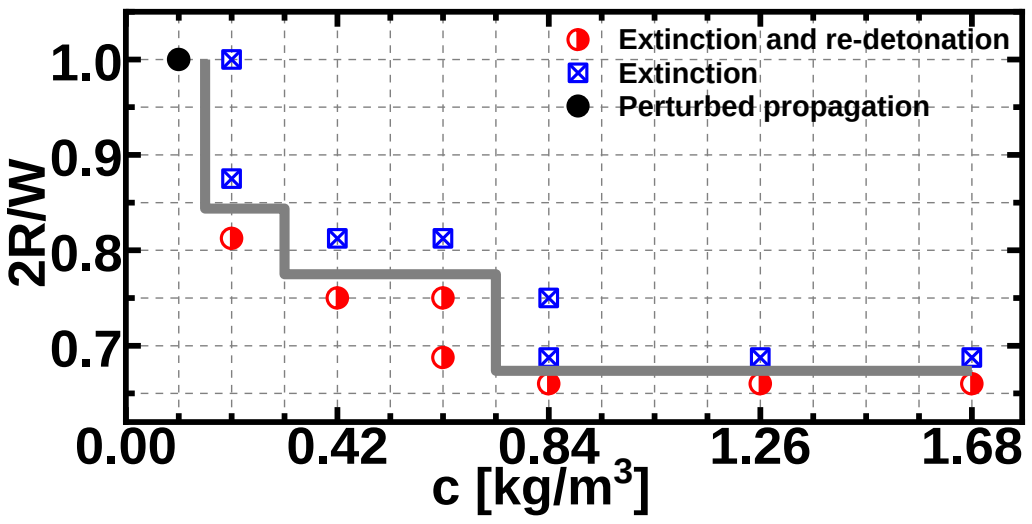

Fig. 11. Diagram of DW propagation and extinction.

Figure 11 demonstrates a diagram of DW propagation and extinction subject to a water cloud. The droplet diameter is fixed to be 2.5 μm, but various droplet concentrations ($c$ =0.105─1.68 kg/m$^3$) and cloud radii ($R$=0.25─0.5$W$) are considered. Parametric studies are performed, and the curve of critical cloud size is determined in Fig. 11. It is observed that the critical water cloud size 2$R$ decreases when the droplet concentration is increased. Below or left to this curve, perturbed propagation (for relatively small $c$) or re-detonation occurs, whilst above it the incident detonation wave is quenched without re-initiation. When $c$ is larger than 0.84 kg/m$^3$, the critical cloud size is negligibly influenced by the droplet concentration.

## 5. Conclusions

Interactions between a propagating detonation wave and water spray cloud in stoichiometric hydrogen/air mixture are simulated by Eulerian − Lagrangian approach and detailed chemical mechanism. A range of droplet diameter, concentration and cloud size are considered.

It is seen that droplet size, concentration and cloud radius have significant effects on peak pressure trajectories of the detonation wave. Three detonation propagation modes are observed: perturbed propagation, leeward re-detonation, and extinction. Leeward re-detonation is analyzed with unsteady evolutions of gas phase and liquid phase quantities. The refracted detonation wave inside the cloud is decoupled and propagates more slowly than the outer detonation. The re-initiation is caused by the shock focusing from upper and lower diffracted detonations.

Breakup of water droplets proceeds when the detonation wave crosses the cloud and multiphase instability of the cloud boundary is observed due to the difference in effective density of the two fluids.

Furthermore, detonation extinction is observed when we vary the water cloud size. It is featured by quickly fading peak pressure trajectories when the detonation passes the cloud, and no local autoignition occurs in the shock focusing area. Evolutions of thermochemical structures along the domain centerline between the reaction front and shock front in an extinction process are also studied.

A series of cloud-gas interaction simulations considering various droplet concentrations and cloud radii are performed. It is shown that the critical cloud size to quench a detonation decreases when the droplet concentration is increased. However, when the droplet concentration is beyond 0.84 kg/m$^3$, this critical cloud size is almost not affected by it.

## Acknowledgements

This work is funded by National Research Foundation (R-265-000-A57-592). YX is supported by the NUS Research Scholarship.